\documentclass[10pt,a4paper,twocolumn]{article}
\usepackage{graphicx}
\usepackage[explicit]{titlesec}
\usepackage{authblk}
\usepackage{etoolbox}
\usepackage[labelfont=bf,labelsep=endash]{caption}
\usepackage{balance}
\usepackage{tabu}
\usepackage{xcolor}
\usepackage[hidelinks]{hyperref}
\usepackage[hang]{footmisc}
\usepackage[normalem]{ulem}
\usepackage[top=1in, bottom=1in, left=0.69in, right=0.69in]{geometry}
\usepackage[T1]{fontenc}
\usepackage{newtxmath,newtxtext}
\usepackage{mathtools}
\usepackage[none]{hyphenat}
\usepackage[numbers]{natbib} 

\usepackage{tabularx} 
\usepackage{xurl} 

\providecommand{\keywords}[1]{\textbf{Keywords} - #1}

\titleformat{\section}{\center\normalfont\bfseries}{\thesection.}{1em}{\MakeUppercase{#1}}
\titlespacing*{\section}{0pt}{12pt}{9pt}

\titleformat{\subsection}{\normalfont\bfseries}{\thesubsection}{1em}{#1}
\titlespacing*{\subsection}{0pt}{12pt}{9pt}

\titleformat{\subsubsection}{\normalfont\itshape}{\thesubsubsection}{1em}{#1}
\titlespacing*{\subsubsection}{0pt}{12pt}{9pt}

\newcommand{\ITUurl}[1]{\textcolor{blue}{\urlstyle{same}\url{#1}}}

\newcommand{\ITUpar}{\vspace{10pt}\par}

\def\starttable{\vspace{6pt}\begin{table}[ht]\center}
\def\startfigure{\vspace{6pt}\begin{figure}[ht]\center}

\makeatletter
\def\tagform@#1{\maketag@@@{\ignorespaces#1\unskip\@@italiccorr}}
\makeatother

\title{\large{\textbf{\uppercase{Advancing Trustworthy AI for Sustainable Development: Recommendations for Standardising AI Incident Reporting}}}}

\author[1]{\normalsize{\textit{Avinash, Agarwal}}}
\author[2]{\normalsize{\textit{Manisha, Nene}}}

\affil[1]{\normalsize{(avinash.70@gov.in) Telecommunication Engineering Centre, Ministry of Communications, New Delhi, India}}
\affil[2]{\normalsize{(mjnene@diat.ac.in) Defence Institute of Advanced Technology, Ministry of Defence, Pune, India}}

\date{}
\begin{document}

\maketitle
\thispagestyle{empty}
\begin{abstract}
\textit{The increasing use of AI technologies has led to increasing AI incidents, posing risks and causing harm to individuals, organizations, and society. This study recognizes and addresses the lack of standardized protocols for reliably and comprehensively gathering such incident data crucial for preventing future incidents and developing mitigating strategies. Specifically, this study analyses existing open-access AI-incident databases through a systematic methodology and identifies nine gaps in current AI incident reporting practices. Further, it proposes nine actionable recommendations to enhance standardization efforts to address these gaps. Ensuring the trustworthiness of enabling technologies such as AI is necessary for sustainable digital transformation. Our research promotes the development of standards to prevent future AI incidents and promote trustworthy AI, thus facilitating achieving the UN sustainable development goals. Through international cooperation, stakeholders can unlock the transformative potential of AI, enabling a sustainable and inclusive future for all.}
\end{abstract}

\begin{center}
\end{center}

\section{Introduction} 
\label{sec:sec1}
The proliferation of AI technologies across diverse domains has led to rapidly increasing AI incidents ranging from algorithmic biases and deepfakes to system failures and unintended consequences. These incidents pose risks to individuals, organizations, and society, thus undermining overall trust and confidence in AI technologies. Recognizing the importance of addressing these risks, stakeholders are increasingly focusing on identifying, analyzing, and mitigating AI-related risks and harms.\ITUpar

Sustainable digital transformation, driven by innovative technologies such as AI, can accelerate progress towards the United Nations' Sustainable Development Goals (SDGs), for example by enhancing access to quality healthcare (SDG 3) \cite{schwalbe2020artificial}, education (SDG 4) \cite{kabudi2022artificial}, managing water crisis and sanitation (SDG 6) \cite{goralski2020artificial}, and climate change adaptation (SDG 13) \cite{leal2022deploying}, among other benefits. However, realizing this potential requires a concerted effort to ensure that AI technologies are deployed responsibly and ethically, with due consideration for their societal and environmental impacts. Several studies highlight that if not deployed responsibly and ethically, AI could impede the achievement of the UN SDGs \cite{vinuesa2020role, gupta2021assessing}. For instance, algorithmic biases in hiring processes could exacerbate inequalities in employment opportunities, thereby impeding progress toward SDG 8 (Decent Work and Economic Growth) and SDG 10 (Reduced Inequalities) \cite{gomez2022gender}.\ITUpar

Principles of Responsible AI, such as those proposed by Organization for Economic Co-operation and Development (OECD), emphasize inclusive growth, sustainability, fairness, transparency, robustness, and accountability of AI systems \cite{OECDRAIprinc}. Compliance with these principles can ensure that AI systems aid and do not hamper the achievement of UN SDGs. Standards, benchmarks, and standardized assessment procedures are needed to ensure that AI systems meet the responsible AI principles \cite{truby2020governing, agarwal2023seven}. Comprehensive data collected through different AI lifecycle stages and deployments in diverse scenarios drives the assessment of compliance with these responsible AI principles.\ITUpar

Learning from past AI incidents is a crucial way to avoid repeat incidents. The aviation industry has well-established protocols for collecting aviation incident-related data. Systematically collecting and analyzing details of aviation incidents have resulted in continuous product improvement and mitigation strategies, leading to a drastic reduction in aviation accidents \cite{gao2021dynamics, dong2021identifying}. Similarly, cybersecurity incident reporting is well-established and supported by regulations in many countries \cite{schmitz2023defining}.\ITUpar

Transparent disclosure of incidents, comprehensive compilation of AI incident data, and their systematic analysis can provide crucial data for developing mitigation strategies and promoting the deployment of trustworthy AI \cite{avin2021filling}. Against the backdrop of the UN SDGs, which seek to address pressing global challenges and promote sustainable development, the need for standardized AI incident reporting becomes even more pronounced. This paper identifies and addresses the critical gap in the availability of standards and protocols for systematic AI incident reporting and data sharing.\ITUpar

In light of these considerations, this paper explores the intersection of AI incident reporting, sustainable digital transformation, and the UN SDGs. By analyzing the current state of AI incident reporting, identifying gaps and challenges, and proposing recommendations for improvement, this research seeks to advance our understanding of how standardization efforts can contribute to achieving sustainable development goals while mitigating AI-related risks. Through collaborative efforts and international cooperation, stakeholders can harness the transformative potential of AI to create a more sustainable and inclusive future for all.\ITUpar

Specific contributions of this study include:
\begin{enumerate}
  \item It identifies nine gaps in existing AI incident reporting practices, offering insights into areas for improvement.
  \item It proposes nine actionable recommendations to enhance standardization efforts in AI incident reporting, addressing the identified gaps.
  \item It facilitates the development of strategies and mechanisms to prevent similar incidents from occurring in the future, thereby promoting trustworthy AI and aligning with the UN SDGs.
\end{enumerate}
The paper is structured as follows: Section 2 reviews the existing literature, delves into the definitions of AI incidents, and reviews available AI incident repositories. Section 3 elaborates on the methodology employed in this study. Observations and results are presented in Section 4, while Section 5 analyses these observations, identifies gaps, draws inferences, and offers corresponding recommendations. Finally, Section 6 provides a summary of the recommendations and conclusions drawn.\ITUpar

\section{Literature Review}
\label{sec:sec2}
\subsection{AI incident definitions}
\label{sec:sec2.1}
The review shows that multiple definitions of “AI incident" are available.\ITUpar
OECD \cite{OECDAIIncident} defines an “AI incident” as, “\emph{an event where the development or use of an AI system: (i) caused harm to person(s), property, or the environment; or (ii) infringed upon human rights, including privacy and non-discrimination}”.\ITUpar
According to the AI Incident Database (AIID), an “AI incident” is “\emph{an alleged harm or near harm event to people, property, or the environment where an AI system is implicated}” \cite{AIIncidentDatabase}.\ITUpar
‘AI, Algorithmic, and Automation Incidents and Controversies’ (AIAAIC) considers an “incident” in the context of AI as “\emph{a sudden known or unknown event (or ‘trigger’) that becomes public and which takes the form of a disruption, loss, emergency, or crisis}” \cite{AIAAICRepository}.\ITUpar
The review reveals the gap related to a lack of standard terms, definitions, and taxonomies. 
\subsection{The need for AI incident reporting}
\label{sec:sec2.2}
Recording AI incidents is crucial for understanding their impact on people, infrastructure, and technology, allowing the development of flexible regulations that evolve with new information and ensure the safe and effective use of AI technologies \cite{lupo2023risky}. Sharing AI incidents improves the verifiability of claims in AI development, highlights overlooked risks, and enhances the effectiveness of external scrutiny by increasing common knowledge of potential AI system behaviors \cite{brundage2020toward}. AI community is starting to recognize incident sharing as vital to prevent vulnerabilities, biases, and privacy concerns in AI systems, ensuring their trustworthiness and enhancing user experience \cite{li2023trustworthy}. Public databases cataloging global AI incidents promote awareness of potential AI harms among policymakers, researchers, and the public, essential for developing safe AI systems \cite{turri2023we}. Collecting real-world failures in incident databases, such as those in mature industrial sectors like aviation, is crucial for informing safety improvements and preventing repeated mistakes in designing and deploying intelligent systems \cite{mcgregor2021preventing}. The collected AI incident data highlights unethical AI use, with top-ranking applications including language and computer vision models, intelligent robots, and autonomous driving, revealing issues like misuse, racism, and bias \cite{nasim2022artificial}.
\subsection{AI incident repositories}
\label{sec:sec2.3}
The AI Incident Database (AIID) \cite{AIIncidentDatabase} is among the earliest initiatives solely focused on documenting AI incidents. It compiles real-world harms or near harms caused by AI systems. Inspired by similar databases in aviation and cybersecurity, AIID aims to draw insights from past incidents to prevent or minimize future adverse outcomes. Another notable repository is the AIAAIC Repository \cite{AIAAICRepository}, which compiles incidents and controversies driven by and relating to AI, algorithms, and automation. The AI Vulnerability Database (AVID) \cite{AVIdatabase} is an open-source repository that aims to catalog failure modes for AI models, datasets, and systems. Its objectives include constructing a comprehensive taxonomy of potential AI harms spanning security, ethics, and performance dimensions and storing detailed information on evaluation use cases and mitigation techniques for each harm category. Another database, the AI Litigation Database (AILD) \cite{AILdatabase} compiles ongoing and completed legal cases concerning artificial intelligence, machine learning, and related fields, offering comprehensive coverage from complaints to verdicts. Further, the OECD.AI expert group is developing the AI Incidents Monitor (AIM) \cite{OECDAIM} to track real-time AI incidents for informing policy discussions. Unlike AIID and AIAAIC, AIM currently does not accept open submissions.\ITUpar
Existing AI incident repositories rely on media coverage and voluntary public submissions, lacking robust mechanisms for technical input \cite{lupo2023risky}. Taxonomies prioritize policy and ethics over technical details, while definitions of AI incidents remain inconsistent \cite{turri2023we}. Moreover, there is a notable absence of federally operated databases, leaving incident reporting reliant on public sources and lacking mandatory legal disclosure and validation processes \cite{turri2023we, agarwal2024addressing}.

\section{Methodology}
\label{sec:sec3}
The study adopted the following methodology:
\begin{enumerate}
\item Executed an exhaustive search and literature review to discover AI incident repositories.
\item Isolated four potential repositories: AIAAIC, AIID, AILD, and AVID. Given that AILD focuses on AI related legal aspects and AVID emphasizes identifying AI system vulnerabilities, shortlisted the two open-access repositories, AIID and AIAAIC, for further scrutiny.
\item Examined the policies, scope, reporting procedures, and review mechanisms of the AIID and AIAAIC databases to comprehend their operational frameworks.
\item Submitted an incident to each database to discern their reporting protocols and procedural intricacies.
\item Retrieved and scrutinized publicly available data from both databases to evaluate their content and structure.
\item Investigated the repositories to pinpoint gaps in standardization across various dimensions, including: incident reporting protocols, quality control, data interoperability, comprehensiveness of data, contributor and source diversity, sector-specific coverage, geographical coverage, and data sharing protocols.
\item Tabulated observations and inferred key insights based on the conducted analysis.
\item Formulated recommendations for standardization activities to address identified gaps and enhance the effectiveness of AI incident reporting practices.
\end{enumerate}

\section{Results}
\label{sec:sec4}
This section presents the observations and results of the study. The next section analyses and draws inferences from them.
\subsection{Incident reporting}
\label{sec:sec4.1}
Table \ref{table1} provides the basics of incident reporting in AIAAIC and AIID. Both have similar processes for incident reporting, though their scopes are slightly different.
\begin{table}[h]
\caption{Incident reporting in AIAAIC and AIID}
\label{table1}
\begin{center}
\begin{tabularx} {\columnwidth} {|
>{\raggedright\arraybackslash}>{\hsize=1.4\hsize} X |
>{\raggedright\arraybackslash}>{\hsize=0.8\hsize} X |
>{\raggedright\arraybackslash}>{\hsize=0.8\hsize} X |
}
\hline
                                        & \textbf{AIAAIC}    & \textbf{AIID}      \\ \hline
What can be reported & Incidents and controversies driven by and relating to AI. & Real-world harms or near harms caused by AI systems. \\ \hline
Incidents reported (as on 05-05-2024)   & 905       & 657       \\ \hline
Who can report incidents                & Anyone    & Anyone    \\ \hline
Submissions reviewed before publishing? & Yes       & Yes       \\ \hline
Nature of reporting                     & Voluntary & Voluntary \\ \hline
Incentive for reporting                 & None      & None      \\ \hline
\end{tabularx}
\end{center}
\end{table}
\subsection{AI-Incident snapshot}
\label{sec:sec4.2}
Sample incidents reported in AIAAIC, shortlisted for analysis, are listed in Table \ref{table2}. These were extracted for analysis by filtering on the criteria “Occurred” = “2024” and “Country(ies)” = “Global”.
\begin{table}[h]
\caption{Snapshot of Incidents reported in AIAAIC}
\label{table2}
\setlength{\tabcolsep}{2pt}  
\begin{center}
\begin{tabularx} {\columnwidth} {|
>{\raggedright\arraybackslash}>{\hsize=0.8\hsize} X |
>{\raggedright\arraybackslash}>{\hsize=1.9\hsize} X |
>{\centering\arraybackslash}>{\hsize=0.3\hsize} X |
}
\hline
\textbf{AIAAIC ID\#} & \textbf{Headline}                                        & \textbf{Ref.}  \\ \hline
AIAAIC1449 & Adobe trained Firefly AI model on competitor images                & {\cite{AIAAIC1449}} \\ \hline
AIAAIC1439 & OpenAI scrapes YouTube to train GPT-4                              & {\cite{AIAAIC1439}} \\ \hline
AIAAIC1414 & Leonardo AI generates celebrity non-consensual porn images         & {\cite{AIAAIC1414}} \\ \hline
AIAAIC1395 & Scientific journals publish papers with AI-generated introductions & {\cite{AIAAIC1395}} \\ \hline
AIAAIC1368 & Microsoft Copilot generates fake Putin comments on Navalny death   & {\cite{AIAAIC1368}} \\ \hline
AIAAIC1356 & ChatGPT 'goes crazy', speaks gibberish                             & {\cite{AIAAIC1356}} \\ \hline
\end{tabularx}
\end{center}
\end{table}
\subsection{Interoperability and data sharing}
\label{sec:sec4.3}
Table \ref{table3} compares the data fields available in the two databases. They have different data structures.
\begin{table}[h]
\caption{Comparison of data fields available in AIID and AIAAIC 
}
\label{table3}
\setlength{\tabcolsep}{2pt}  
\begin{center}
\begin{tabularx} {\columnwidth} {|
>{\raggedright\arraybackslash}>{\hsize=1\hsize} X |
>{\raggedright\arraybackslash}>{\hsize=1\hsize} X |
>{\raggedright\arraybackslash}>{\hsize=1\hsize} X |
}
\hline
\textbf{Fields available in both AIID and AIAAIC} &
  \textbf{Fields available only in AIID} &
  \textbf{Fields available only in AIAAIC} \\ \hline
\begin{tabular}[X]{@{}X@{}}Incident ID;\\ Title/ Headline;\\ Description;\\ Occurrence date;\\System deployer;\\ System developer;\end{tabular} &
  Alleged harmed or nearly or nearly harmed parties &
  \begin{tabular}[X]{@{}X@{}}Type;\\ Released (year);\\ Country(ies);\\ Sector(s);\\ System name(s);\\ Technology(ies);\\ Purpose(s);\\ Media trigger(s);\\ Issue(s);\\ Transparency;\\External harms;\\ Internal harms\\ \end{tabular} \\ \hline
\end{tabularx}
\end{center}
\end{table}
\subsection{Contributors to the Databases}
\label{sec:sec4.4}
Table \ref{table4} lists the top seven submitters of the published incidents in AIID. They reported more than 70{\%} of all the incidents in AIID. AIAAIC does not have data fields to capture this data.
\begin{table}[h]
\caption{Top seven submitters of the incidents in AIID}
\label{table4}
\begin{center}
\begin{tabularx} {\columnwidth} {|
>{\raggedright\arraybackslash}>{\hsize=1.4\hsize} X |
>{\centering\arraybackslash}>{\hsize=0.8\hsize} X |
>{\centering\arraybackslash}>{\hsize=0.8\hsize} X |}
\hline

\textbf{Submitters}     & \textbf{Incidents} & \textbf{\%age} \\ \hline
Daniel Atherton         & 149                & 23\%           \\ \hline
Anonymous               & 96                 & 15\%           \\ \hline
Khoa Lam                & 93                 & 14\%           \\ \hline
Ingrid Dickinson CSET & 49                 & 7\%            \\ \hline
Roman Yampolskiy        & 29                 & 4\%            \\ \hline
AIAAIC                  & 25                 & 4\%            \\ \hline
Kate Perkins            & 21                 & 3\%            \\ \hline
\end{tabularx}
\end{center}
\end{table}
\subsection{Sources of the reports submitted to the databases}
\label{sec:sec4.5}
Table \ref{table5} provides details of the top seven source domains of the reports submitted to AIID. AIAAIC does not have data fields to capture this data.
\begin{table}[h]
\caption{Top seven source-domains of the reports in AIID}
\label{table5}
\begin{center}
\begin{tabularx} {\columnwidth} {|
>{\centering\arraybackslash}>{\hsize=1\hsize} X |
>{\centering\arraybackslash}>{\hsize=1\hsize} X |}
\hline
\textbf{Source domain}      & \textbf{Reports} \\ \hline
theguardian.com    & 143     \\ \hline
theverge.com       & 95      \\ \hline
nytimes.com        & 94      \\ \hline
washingtonpost.com & 71      \\ \hline
wired.com          & 69      \\ \hline
vice.com           & 54      \\ \hline
reuters.com        & 53      \\ \hline
bbc.com            & 53      \\ \hline
\end{tabularx}
\end{center}
\end{table}
\subsection{Sector Coverage}
\label{sec:sec4.6}
Table \ref{table6} details the top seven sectors of the incidents reported in AIAAIC. While AIID does not have data fields to capture this data, Table \ref{table7} provides details of the top seven deployers of the AI systems with incidents reported in AIID.
\begin{table}[h]
\caption{Top seven sectors of the incidents in AIAAIC}
\label{table6}
\begin{center}
\begin{tabularx} {\columnwidth} {|
>{\raggedright\arraybackslash}>{\hsize=1.9\hsize} X |
>{\centering\arraybackslash}>{\hsize=0.55\hsize} X |
>{\centering\arraybackslash}>{\hsize=0.55\hsize} X |}
\hline
\textbf{Sectors}                & \textbf{Incidents} & \textbf{\%age} \\ \hline
Media/entertainment/sports/arts & 193                         & 21.3\%         \\ \hline
Automotive                      & 86                          & 9.5\%          \\ \hline
Politics                        & 75                          & 8.3\%          \\ \hline
Technology                      & 60                          & 6.6\%          \\ \hline
Education                       & 58                          & 6.4\%          \\ \hline
Banking/financial services      & 40                          & 4.4\%          \\ \hline
Business/professional services  & 35                          & 3.9\%          \\ \hline
\end{tabularx}
\end{center}
\end{table}
\begin{table}[h]
\caption{Top seven deployers of the AI systems in AIID}
\label{table7}
\begin{center}
\begin{tabularx} {\columnwidth} {|
>{\raggedright\arraybackslash}>{\hsize=1.4\hsize} X |
>{\centering\arraybackslash}>{\hsize=0.8\hsize} X |
>{\centering\arraybackslash}>{\hsize=0.8\hsize} X |}
\hline
\textbf{Deployer of AI system} & \textbf{incidents} & \textbf{\%age} \\ \hline
tesla                          & 39                 & 6\%            \\ \hline
facebook                       & 36                 & 6\%            \\ \hline
google                         & 28                 & 4\%            \\ \hline
unknown                        & 23                 & 4\%            \\ \hline
amazon                         & 21                 & 3\%            \\ \hline
openai                         & 20                 & 3\%            \\ \hline
cruise                         & 12                 & 2\%            \\ \hline
\end{tabularx}
\end{center}
\end{table}
\subsection{Geographical coverage}
\label{sec:sec4.7}
Table \ref{table8} lists the top seven countries related to the geographic origin and/or primary extent of the incidents reported in AIAAIC. While AIID does not have data fields to capture this data, as indicated in Table \ref{table7}, the incidents reported in AIID are predominantly related to AI systems developed by American companies.
\begin{table}[h]
\caption{Top seven countries of the incidents in AIAAIC}
\label{table8}
\begin{center}
\begin{tabularx} {\columnwidth} {|
>{\raggedright\arraybackslash}>{\hsize=1.4\hsize} X |
>{\centering\arraybackslash}>{\hsize=0.8\hsize} X |
>{\centering\arraybackslash}>{\hsize=0.8\hsize} X |}
\hline
\textbf{Countries} & \textbf{Incidents} & \textbf{\%age} \\ \hline
USA                & 424                & 46.9\%         \\ \hline
UK                 & 59                 & 6.5\%          \\ \hline
China              & 53                 & 5.9\%          \\ \hline
USA; Global        & 26                 & 2.9\%          \\ \hline
Global             & 21                 & 2.3\%          \\ \hline
India              & 21                 & 2.3\%          \\ \hline
Canada             & 18                 & 2.0\%          \\ \hline
\end{tabularx}
\end{center}
\end{table}
\subsection{Data sharing}
\label{sec:sec4.8}
Table \ref{table9} outlines the formats available for downloading incident data from the two databases and the limitations on accessible data.
\begin{table}[h]
\caption{Sharing of incident data by AIAAIC and AIID}
\label{table9}
\begin{center}
\begin{tabularx} {\columnwidth} {|
>{\raggedright\arraybackslash}>{\hsize=0.7\hsize} X |
>{\centering\arraybackslash}>{\hsize=1.15\hsize} X |
>{\centering\arraybackslash}>{\hsize=1.15\hsize} X |}
\hline
\textbf{Data sharing} & \textbf{AIAAIC} & \textbf{AIID} \\ \hline
Format &
  Available as a Google Sheet. &
  Weekly snapshots of the database in JSON, MongoDB, and CSV format \\ \hline
Information not accessible &
  Contributor details are not public. Harm data is only accessible to premium members. &
  - \\ \hline
APIs                  & None            & None          \\ \hline
\end{tabularx}
\end{center}
\end{table}

\section{Gap analysis and recommendations}
\label{sec:sec5}
This section analyses the results to identify gaps in existing AI-incident reporting mechanisms and recommends areas for standardization and policy initiatives. These recommendations aim to address observed gaps, enabling meticulous AI-incident reporting and contributing to the achievement of the UN SDGs.
\subsection{Lack of definitions and taxonomies}
\label{sec:sec5.1} 
\textbf{Observation:} There is a lack of consistency in qualifying the reported events as incidents. The AIAAIC incidents with ids AIAAIC1449 \cite{AIAAIC1449} and AIAAIC1439 \cite{AIAAIC1439} cited in Table \ref{table2} relate to ethical practices and possible copyright infringement, but qualifying them as AI-incidents will depend on the definition of AI-incident. Similarly, incident id AIAAIC1395 \cite{AIAAIC1395} at s.no. 4 in Table (\ref{table2}) relates to the ethics of the authors and the screening processes followed by the journals and does not meet the AI-incident definition provided by OECD \cite{OECDAIIncident}. Also, it is challenging to determine the severity of the incidents based on the information available in both databases.\ITUpar

\textbf{Inference:} One significant gap is the absence of standardized definitions and taxonomies related to AI incidents and AI harms. It becomes challenging to compare and analyze incidents across different domains and jurisdictions without consistent guidelines for categorizing incidents, their harms, and severity levels.\ITUpar

\textbf{Recommendation 1:}\emph{Standardise AI-incident and AI-harms taxonomies:} Develop standard taxonomies for AI-incidents and AI-harms based on domain, severity, root causes, and impact on SDGs to enable consistent classification and analysis of AI-incidents across different sectors and jurisdictions, facilitating benchmarking and trend analysis.

\subsection{Bias, inconsistencies, and misclassification}
\label{sec:sec5.2} 
\textbf{Observation:} As mentioned in the previous paragraph, three of the incidents cited in Table \ref{table2}  \cite{AIAAIC1449}, \cite{AIAAIC1439}, and \cite{AIAAIC1395} may not qualify as AI incidents depending on the definition considered. The reporting of incidents, their review, classification as incidents, and assessing their harm quotients being manual are prone to biases and capabilities of the individuals involved. Biases and inconsistencies in incident reporting can skew perceptions of AI-related risks and hinder efforts to develop inclusive and equitable solutions.\ITUpar

\textbf{Inference:} The AI-incident databases may suffer from the biases of the submitters or the reviewers related to attributes such as their political leanings, gender, minority groups, countries, and so on. Further, different individuals classify the incidents and their harms in distinct ways, depending on their exposure, capabilities, and understanding, which may lead to inconsistencies and misclassification.\ITUpar

\textbf{Recommendation 2:} \emph{Define guidelines for AI-incident database quality audits:} Formulate procedures to regularly audit the AI-incident databases for consistency, checking for misreporting, misclassification, reported incidents meeting the defined criteria, and so on.

\subsection{Insufficient and incompatible data fields}
\label{sec:sec5.3} 
\textbf{Observation:} Table \ref{table3} compares the columns available in the two databases, showing that only six fields are compatible between the two datasets, while the remaining are incompatible. Secondly, these databases do not have enough detailed data fields needed for thorough analysis, like identifying the causes, context, and impact of reported incidents. AIID does not have fields to capture impacted sectors (Table \ref{table6}), impacted countries (Table \ref{table8}), and so on. On the other hand, AIAAIC does not capture details of the harmed (or nearly harmed) parties the way AIID does, such as Facebook users, minority groups, patients, and so on.\ITUpar

\textbf{Inference:} Different and incompatible structures of AI incident databases make aggregating data from multiple databases difficult, limit interoperability, and restrict data exchange. Secondly, the captured data is generally insufficient for assessing the severity and proper categorization of the incidents.\ITUpar

\textbf{Recommendation 3:} \emph{Standardise AI-incident database structures:} Standardising the fields of AI-incident databases will ensure that the collected data has sufficient granularity required for analysis. It will also facilitate interoperability, data exchange, and ease of aggregating data from multiple databases.  

\subsection{Inadequate motive to report incidents}
\label{sec:sec5.4} 
\textbf{Observation:} As indicated in Table \ref{table1}, incident reporting in both databases is voluntary and lacks incentives. Without legal mandates or rewards, reporting relies on reporters' discretion and motivation, potentially resulting in underreporting. \ITUpar

\textbf{Inference:} Fears of data privacy breaches may discourage reporting, leading to incomplete or underreported AI incidents. Without transparent and privacy-protective reporting mechanisms, stakeholders may hesitate to disclose incidents, hampering the effectiveness of incident databases. Additionally, fragmentation among databases complicates data collection and analysis, impeding comprehensive risk understanding and response.\ITUpar

\textbf{Recommendation 4:} \emph{Develop regulatory and policy frameworks for AI-incident reporting:} Make sector-specific legal provisions to mandate or encourage AI-incident reporting. Global standards organizations such as ITU should develop standardized regulatory and policy frameworks for AI-incident reporting to enable consistency across nations.

\subsection{Narrow base of the incidents reported}
\label{sec:sec5.5} 
\textbf{Observation:} Though the incident reporting is open to the public, only a few individuals report the incidents. Table \ref{table4} indicates that just four individuals, excluding the anonymous ones, have reported half of the incidents in AIID. Further, the top sources of the reports submitted to AIID are from American or European newspapers, as detailed in Table \ref{table5}.\ITUpar

\textbf{Inference:} Technological interventions and process reforms are required to widen the base of incident reporting.\ITUpar

\textbf{Recommendation 5:} \emph{Develop standards for automated incident reporting:} Develop standards to enable automated AI-incident reporting through the AI applications to supplement manual reporting.

\subsection{Inadequate data-sharing protocols}
\label{sec:sec5.6} 
\textbf{Observation:} As indicated in Table \ref{table9}, the two databases allow downloading data in different formats, and both do not provide APIs for accessing data. Further, there is inconsistency related to the information accessible from the two databases (Table \ref{table9}). The submitter names are accessible in AIID but not in AIAAIC. Similarly, AIID provides access to the details of the harmed parties, but in AIAAIC, harm data is only accessible to Premium Members.\ITUpar

\textbf{Inference:} Therefore, standardized mechanisms for sharing incident data among stakeholders, including government agencies, industry partners, researchers, and the public, are lacking. It impedes collaborative efforts to address emerging trends, root causes, and mitigation strategies for AI incidents.\ITUpar

\textbf{Recommendation 6:} \emph{Standardise data sharing mechanisms:} Define protocols for data sharing, access controls, and privacy protection to ensure the confidentiality and security of incident data. Establish mechanisms for sharing incident data among stakeholders, including government agencies, industry partners, research institutions, and civil society organizations.

\subsection{Sectoral underrepresentation:}
\label{sec:sec5.7} 
\textbf{Observation:} Existing AI-incident databases have skewed representations of application sectors. "Media/entertainment/sports/art" sector has the highest number of incidents reported in AIAAIC, followed by automotive and politics sectors, as illustrated in Table \ref{table6}. Table \ref{table7} indicates that the maximum incidents reported in AIID relate to self-driving cars (Tesla, Cruise), social media (Facebook), search engines (Google), online shopping (Amazon), and advanced AI models (OpenAI). \ITUpar

\textbf{Inference:} While these databases predominantly report consumer-oriented sectors, they underrepresent critical infrastructure sectors such as telecom and electricity supply. The AI incidents in such sectors may not be as frequent as in the consumer-oriented sectors; however, it is still vital to maintain a repository of their incidents. \ITUpar

\textbf{Recommendation 7:} \emph{Sector-specific AI-incident databases:} Develop sector-specific AI-incident databases to supplement the general purpose AI-incident databases.

\subsection{Demographic underrepresentation:}
\label{sec:sec5.8}
\textbf{Observation:} Table \ref{table8} shows that just three countries account for 60\% of the incidents reported in AIAAIC. Similarly, the incidents reported in AIID predominantly relate to AI systems developed by American companies, as evident from Table \ref{table7}. Further, the top sources of the reports submitted to AIID are from American or European newspapers, as detailed in Table \ref{table5}. \ITUpar

\textbf{Inference:} Existing AI-incident databases particularly lack representation from developing and underdeveloped countries. Capturing AI incidents prevalent in these underrepresented regions is crucial for developing mitigation strategies. It is also essential in advancing the UN SDGs. \ITUpar

\textbf{Recommendation 8:} \emph{ITU-led inclusive AI incident reporting:} Encourage international collaboration facilitated by UN organizations, such as ITU, to establish standardized protocols for AI-incident reporting, prioritizing inclusivity from developing countries. This promotes comprehensive understanding and mitigation aligned with UN SDGs.

\subsection{Lack of awareness:}
\label{sec:sec5.9}
\textbf{Observation:} As mentioned in the previous paragraphs and observed through Tables \ref{table4} and \ref{table5}, the base of AI incident reporting is narrow.\ITUpar

\textbf{Inference:} The key stakeholders, including industry, academia, civil society, the general public, and policymakers, are largely unaware of AI-incident databases. Without active involvement from diverse perspectives, databases will fail to capture the full spectrum of AI-related risks and opportunities. \ITUpar

\textbf{Recommendation 9:} \emph{Awareness programs:} Hold regular campaigns to enhance stakeholders' awareness and understanding of AI incident reporting standards and best practices. \ITUpar

These standardization actions can enhance the effectiveness, transparency, and accountability of AI-incident reporting processes, thereby contributing to the achievement of the UN SDGs. \ITUpar

It is further recommended to include incident reporting as an integral part of the AI lifecycle so that it gets appropriate focus in the future. Figure \ref{fig1} illustrates the conceptualized AI lifecycle stages to collect data for developing incident mitigation strategies.
\startfigure
\includegraphics[width=\columnwidth]{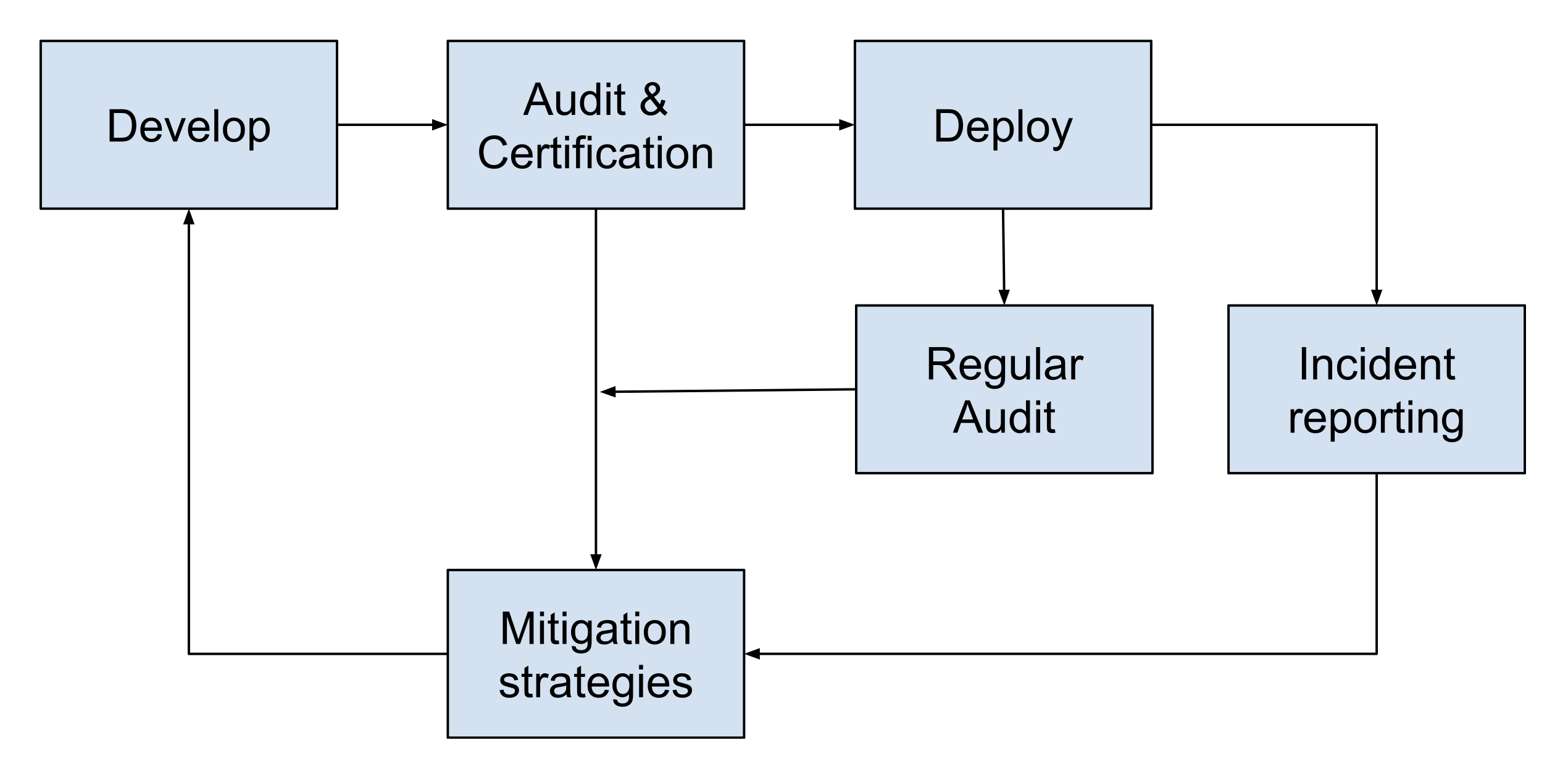}
\caption{Conceptualised AI lifecycle stages}\label{fig1} 
\end{figure}

\section{Conclusion}
\label{sec:sec6}
In conclusion, this study highlights the critical need for standardized AI-incident reporting to enable data gathering, research, and development of mitigation strategies for preventing future incidents. Through an analysis of existing open-access AI-incident databases, it presents the key observations and gaps in standardization, underscoring the importance of policy and standardization initiatives in this domain. Table \ref{table10} summarises the gaps observed and the recommendations to overcome them.\ITUpar

\begin{table}[h]
\caption{Summary of observed gaps and recommendations}
\label{table10}
\setlength{\tabcolsep}{2pt}  
\begin{center}
\begin{tabularx} {\columnwidth} {|
>{\centering\arraybackslash}>{\hsize=0.3\hsize} X |
>{\raggedright\arraybackslash}>{\hsize=1.2\hsize} X |
>{\raggedright\arraybackslash}>{\hsize=1.5\hsize} X |}
\hline
               & \textbf{Gaps observed}                    & \textbf{Recommendations}                        \\ \hline
1              & Lack of definitions and taxonomies        & Standardise AI-incident and AI-harms taxonomies \\ \hline
2 & Bias, inconsistencies, and misclassification           & Define guidelines for AI-incident database quality audits          \\ \hline
3              & Insufficient and incompatible data fields & Standardise AI-incident database structures     \\ \hline
4 & Inadequate motive to report incidents                  & Develop regulatory and policy frameworks for AI-incident reporting \\ \hline
5 & Narrow base of the incidents reported                  & Develop standards for automated incident reporting                 \\ \hline
6              & Inadequate data-sharing protocols         & Standardise data sharing mechanisms             \\ \hline
7              & Sectoral underrepresentation              & Sector-specific AI-incident databases           \\ \hline
8              & Demographic underrepresentation           & ITU-led inclusive AI incident reporting         \\ \hline
9              & Lack of awareness                         & Awareness programs                              \\ \hline
\end{tabularx}
\end{center}
\end{table}

Standardized incident reporting protocols and mechanisms proposed by this study will facilitate data-driven mitigation strategies and product improvement. It will also enable responsible and trustworthy AI deployment for sustainable development. \ITUpar

Overall, the standardization of AI incident reporting is crucial for promoting trust, transparency, and accountability in deploying AI technologies. By implementing the recommendations outlined in this paper, stakeholders can contribute to achieving the UN Sustainable Development Goals and fostering a digital transformation that benefits humanity and the planet. By bridging identified gaps and advancing standardization initiatives, stakeholders can unlock the transformative potential of AI, ushering in a more sustainable and inclusive future for all.\ITUpar

Looking ahead, a concerted effort is required to prioritize multi-stakeholder engagement and international cooperation in standardization endeavors. By harnessing diverse perspectives and expertise, stakeholders can develop robust AI frameworks and guidelines aligned with the tenets of sustainable development, thereby contributing significantly to the attainment of the UN SDGs.\ITUpar

\bibliographystyle{unsrt}
\bibliography{references}

\end{document}